# High-throughput molecular imaging via deep learning enabled Raman spectroscopy

Conor C. Horgan[1,5], Magnus Jensen[1], Anika Nagelkerke[2], Jean-Phillipe St-Pierre[3], Tom Vercauteren[4], Molly M. Stevens[5*], Mads S. Bergholt[1*]

[1]Centre for Craniofacial and Regenerative Biology, King's College London, London SE1 9RT, UK
[2]University of Groningen, Groningen Research Institute of Pharmacy, Pharmaceutical Analysis, P.O. Box 196, XB20, 9700 AD Groningen, The Netherlands
[3]Department of Chemical and Biological Engineering, University of Ottawa, Ottawa, ON K1N 6N5, Canada
[4]School of Biomedical Engineering and Imaging Sciences, King's College London, WC2R 2LS, UK
[5]Department of Materials, Department of Bioengineering, and Institute of Biomedical Engineering, Imperial College London, London SW7 2AZ, UK
*Corresponding authors: mads.bergholt@kcl.ac.uk, m.stevens@imperial.ac.uk

## Abstract

Raman spectroscopy enables non-destructive, label-free imaging with unprecedented molecular contrast but is limited by slow data acquisition, largely preventing high-throughput imaging applications. Here, we present a comprehensive framework for higher-throughput molecular imaging via deep learning enabled Raman spectroscopy, termed DeepeR, trained on a large dataset of hyperspectral Raman images, with over 1.5 million spectra (400 hours of acquisition) in total. We firstly perform denoising and reconstruction of low signal-to-noise ratio Raman molecular signatures via deep learning, with a 9× improvement in mean squared error over state-of-the-art Raman filtering methods. Next, we develop a neural network for robust 2–4× super-resolution of hyperspectral Raman images that preserves molecular cellular information. Combining these approaches, we achieve Raman imaging speed-ups of up to 160×, enabling high resolution, high signal-to-noise ratio cellular imaging in under one minute. Finally, transfer learning is applied to extend DeepeR from cell to tissue-scale imaging. DeepeR provides a foundation that will enable a host of higher-throughput Raman spectroscopy and molecular imaging applications across biomedicine.

## Introduction

Raman spectroscopy has recently excelled as a highly complementary tool for biomedical research providing non-destructive, label-free, molecular imaging with subcellular resolution. This has enabled a multitude of exciting biomedical applications from fundamental *in vitro* cellular studies[1,2], to *ex vivo* spectral histopathology[3,4], and *in vivo* fibreoptic endoscopy for optical biopsy at the molecular level[5,6]. Despite its many advantages, Raman spectroscopy remains limited by the weakness of generated Raman signal, which necessitates spectral acquisition times on the order of one second per sampled point[7]. As such, high-resolution Raman spectroscopic imaging of cells or tissues often requires multiple hours, which is prohibitive for high-throughput Raman spectroscopic imaging applications[8–10]. To address these acquisition time and signal-to-noise ratio (SNR) challenges, advanced nonlinear Raman spectroscopy techniques including coherent anti-stokes Raman spectroscopy (CARS) and stimulated Raman spectroscopy (SRS) have been developed[11–13]. However, while these advanced techniques have recently enabled rapid broadband Raman imaging in biomedicine, they do so through application of pulsed laser systems that are technically demanding to operate and incur significant costs[14].

A potential alternative, or complement, to hardware-based solutions lies in deep learning[15]. Deep learning is a subset of machine learning capable of uncovering effective representations of data across multiple levels of abstraction and has demonstrated incredible results across several domains including image classification and segmentation, natural language processing, and predictive modelling[16–19]. Recently, the application of deep learning to point-based Raman spectroscopy has achieved promising results in the intraoperative diagnosis of brain tumours and in the rapid identification of pathogenic bacteria[20,21]. While such applications are likely to improve Raman spectroscopic diagnostic accuracy, even greater benefits lie in the potential for deep learning to improve Raman spectroscopic imaging by increasing signal acquisition speeds and enabling high-quality reconstruction from noisy, low-resolution input data. Hyperspectral Raman images, where each pixel contains a complete Raman spectrum, represent highly struc-tured data with complex spatial and spectral correlations amenable to deep learning. This highly structured nature of the data is at present underutilised, with existing data processing and analysis techniques (e.g. chemometrics and multivariate analysis) failing to adequately exploit these complex correlations.

Deep learning has made significant strides in signal reconstruction tasks, most notably for single-image super-resolution (SISR)[22–24], image denoising[25,26], and signal denoising[27,28]. In each of these domains, a neural network is trained on pairs of low-quality (e.g. low-resolution (LR)) and high-quality (e.g. high-resolution (HR)) data, attempting to learn effective representations of the low-quality inputs that reconstruct the corresponding high-quality outputs. Neural networks can thus be considered to learn prior information (e.g. shapes, sizes, colours typical of different features) from the corpus of data in the training set in order to generate high-quality output data given low-quality input data in the test set.

In the context of signal denoising, several groups have developed neural networks designed to reduce noise in electrocardiograms[27,28], while image denoising has been employed to improve image quality by removing noise generated by imaging hardware or compression artefacts[26,29]. Similarly, much work has focused on SISR, with important potential life-sciences applications already demonstrated for fluorescence microscopy, MRI, electron microscopy, and even endomicroscopy[30–34]. SISR approaches have enabled HR fluorescence microscopy with ~100× lower light dose and 16× higher frame rates for reduced photo-bleaching and photo-toxicity[32], while 4× brain MRI super-resolution has achieved shorter scan times, increased spatial coverage, and higher signal-to-noise ratio (SNR)[31].

Here, we present DeepeR, a comprehensive deep learning framework for high-throughput molecular imaging via deep learning enabled Raman spectroscopy. We first show, using a dataset consisting of 172,312 pairs of low and high SNR Raman spectra, that deep learning significantly outperforms state-of-the-art spectral smoothing algorithms 9×, enabling effective reconstruction of Raman signatures from low SNR Raman spectra. We next develop a convolutional neural network for hyperspectral Raman image super-resolution, using an additional dataset of 169 hyperspectral images representing 1.4 million Raman spectra (389 hours of acquisition) in total. We achieve robust 2–4× super-resolution image reconstruction, corresponding to a 4–16× reduction in imaging time. Then, using a hybrid approach, we demonstrate Raman imaging with effective speed-ups of 40–160× while preserving molecular cellular information. Finally, we highlight the generalisability of our deep learning framework, employing transfer learning to extend our pre-trained neural networks from cells to tissues.

## Results

### Hyperspectral Raman Deep Learning Framework

DeepeR is designed to improve Raman spectroscopic acquisition times towards high-throughput Raman imaging applications. Working across hyperspectral Raman data, DeepeR performs i) Raman spectral denoising, ii) hyperspectral super-resolution, and iii) transfer learning (Figure 1). Raman spectral denoising is performed using a 1D residual UNet (ResUNet)[35], which takes low SNR input spectra and reconstructs them to produce corresponding high SNR output spectra (Figure 1i). Hyperspectral super-resolution is achieved using an adapted residual channel attention network (RCAN)[36] to output a HR hyperspectral Raman image from a LR input (Figure 1ii, Supplementary Figure 1). The combination of i) Raman spectral denoising and ii) hyperspectral super-resolution then enables significant Raman imaging speed-ups for high-throughput applications. Finally, DeepeR can be generalised to a wide range of Raman imaging applications through transfer learning, where neural networks pre-trained on large hyperspectral datasets can be fine-tuned to operate effectively on small hyperspectral datasets (Figure 1iii).

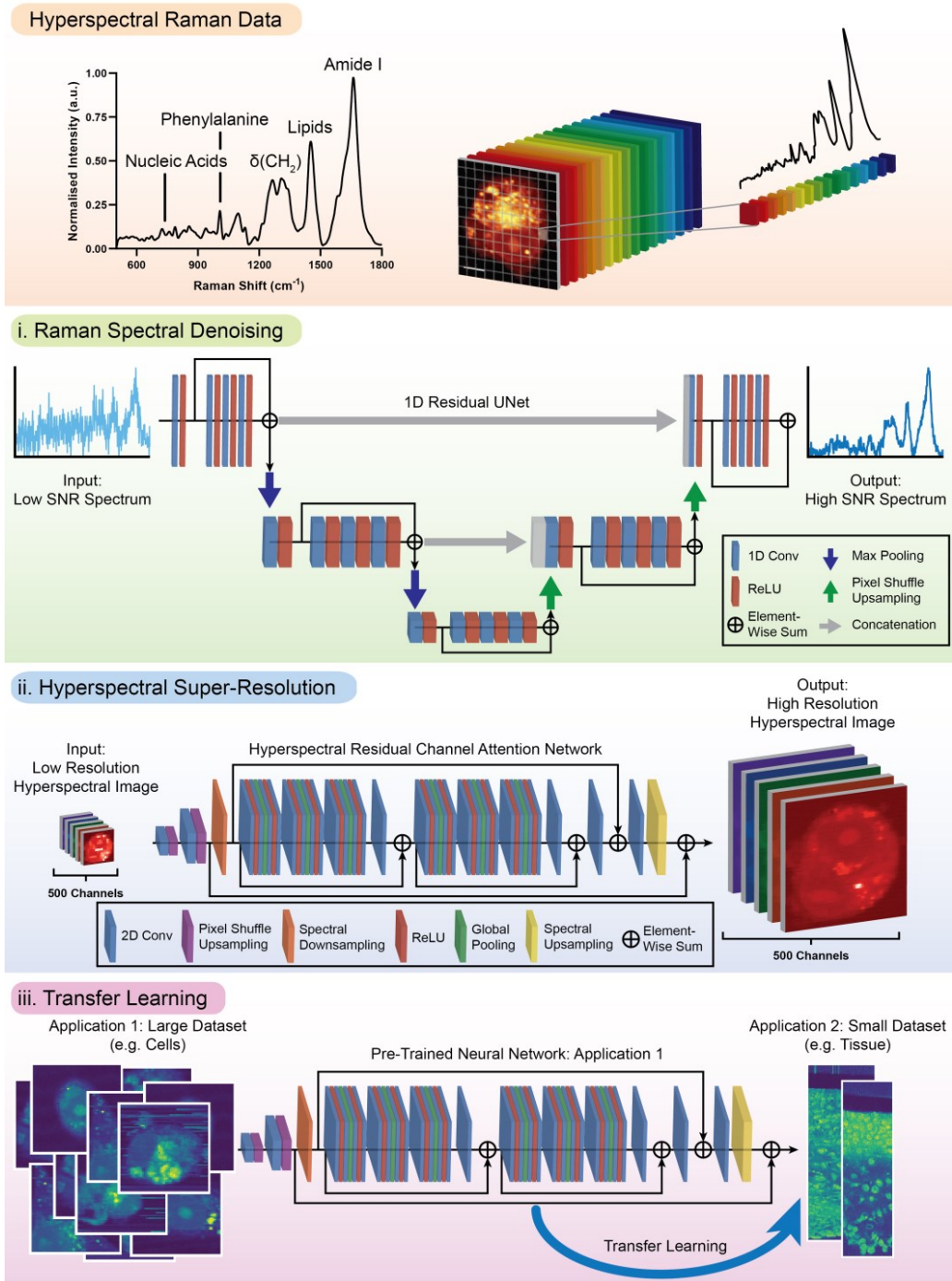

**Figure 1 | Deep Learning Enabled Raman Hyperspectral Super-Resolution Imaging.** The deep learning framework DeepeR is designed to operate on hyperspectral Raman images, where high information-content Raman spectra at each pixel provide detailed insight into the molecular composition of cells/tissues. To improve the speed of Raman spectroscopic imaging and enable high-throughput applications, we first **(i)** train a 1D ResUNet neural network for Raman spectral denoising to effectively reconstruct a high SNR Raman spectrum (long acquisition time) from a corresponding low SNR input spectrum (short acquisition time). Next, we **(ii)** train a hyperspectral residual channel attention neural network to accurately reconstruct high spatial resolution hyperspectral Raman images from corresponding low spatial resolution hyperspectral Raman images to significantly reduce imaging times. Then, by combining **(i)** and **(ii)** we achieve extreme speed-ups of up to 160× in Raman imaging time while maintaining high reconstruction fidelity. Finally, we **(iii)** demonstrate that transfer learning can be used to take our pre-trained neural networks (trained on large datasets) to operate on an entirely unrelated hyperspectral data domain for which there is only a limited dataset (insufficient to effectively train a neural network from scratch).

**Deep Learning Enabled Raman Denoising**

We first developed a neural network training pipeline for Raman denoising – the reconstruction of a Raman spectrum with high SNR from a corresponding low SNR Raman spectrum. We cultured MDA-MB-231 breast cancer cells, a widely studied cell line, and sequentially acquired low SNR (0.1 second integration time per spectrum) and high SNR (1 second integration time per spectrum) hyperspectral confocal Raman cell image pairs (n = 11 cells) using 532 nm laser excitation. This resulted in a large dataset consisting of pairs of low and high SNR Raman spectra (n = 172,312 spectral pairs). Importantly, these Raman spectra contain an abundance of molecular information, including information about the relative concentrations and distributions of various nucleic acids, proteins, and lipids[1]. For instance, intense peaks can be seen near 795 cm$^{-1}$ (DNA), 1004 cm$^{-1}$ (phenylalanine), 1300 and 1440 cm$^{-1}$ (lipids), and 1660 cm$^{-1}$ (predominantly Amide I of proteins). Successful denoising and reconstruction of low SNR Raman spectra requires that this biochemical information be effectively preserved.

We applied a 1D ResUNet to this dataset, performing 11-fold cross-validation by training 11 independent models on training/validation sets composed of all of the spectra from 10 hyperspectral Raman cell images, where the test sets in each case consisted of all the spectra from the remaining hyperspectral Raman cell image (see Materials and Methods and Supplementary Table 1 for full implementation details). The 1D ResUNet learned to produce high quality output Raman spectra from low SNR input spectra that strongly aligned with the target (ground truth) high SNR spectra (Figure 2). To achieve this, the 1D ResUNet operates across multiple feature scales and spectral resolutions, learning spectral features (and the molecular constituents they represent) typical of Raman spectra in the training set in order to identify a mapping between low SNR inputs and target high SNR outputs in the test set. Importantly, the neural network significantly outperformed Savitzky-Golay (SG) filtering (assessed via a two-tailed Wilcoxon paired signed rank test) across multiple SG filter parametrisations, the current state-of-the-art for Raman spectral smoothing[37]. Application of the neural network to the test sets achieved a spectral mean squared error (MSE) between output and target spectra that was 9× lower than the best performing SG filter, with a mean MSE of 2.85x10$^{-3}$ [95% confidence interval (CI): 2.55x10$^{-3}$, 3.15x10$^{-3}$] for the 1D ResUNet and 2.54x10$^{-2}$ [95% CI: 2.30x10$^{-2}$, 2.78x10$^{-2}$] for the best performing SG filter (Figure 2b). This result demonstrates that the 1D ResUNet effectively learned the structure of Raman spectra, enabling it to discern true signal from background noise far better than a naïve SG algorithm (which has no *a priori* knowledge of the typical structures of Raman spectral features).

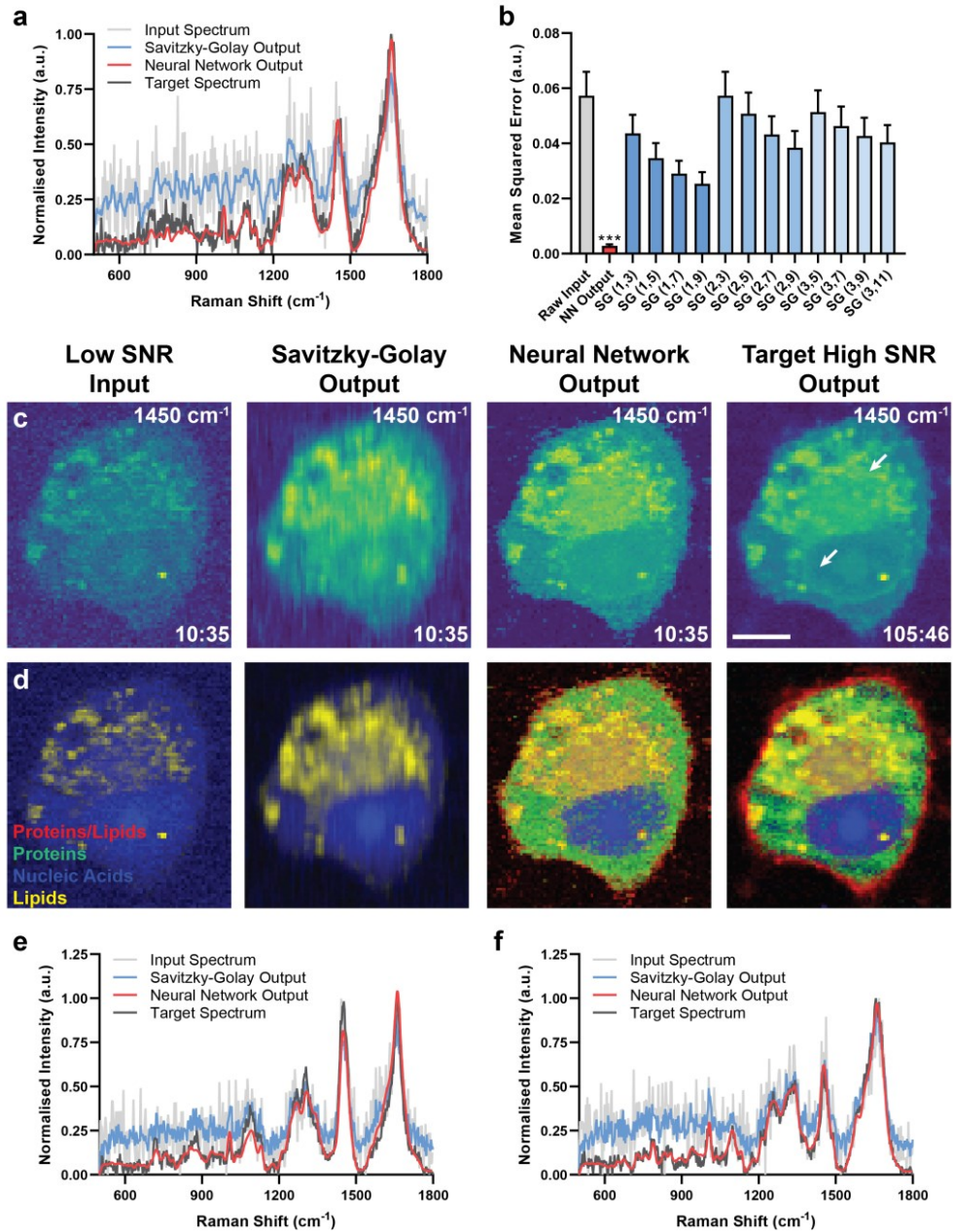

**Figure 2 | Deep Learning Enabled Raman Denoising. (a)** Exemplar test set pair of low SNR input Raman spectrum (light grey) and corresponding high SNR target Raman spectrum (dark grey) as well as the Savitzky-Golay (blue) and neural network (red) outputs for the given input spectrum (normalised to maximum peak intensity). **(b)** Mean squared error (performed across all spectral channels and all image pixels) across all test set hyperspectral Raman cell images for raw input spectra, 1D ResUNet output spectra, and Savitzky-Golay output spectra (order *x*, frame width *y*) with respect to corresponding target spectra (n = 11) (error bars: mean ± STD) (two-tailed Wilcoxon paired signed rank test against best performing Savitzky-Golay filter, *** $P < 0.005$). **(c)** Exemplar 1450 cm$^{-1}$ peak intensity heatmaps for low SNR input hyperspectral Raman image, Savitzky-Golay (1st order, frame length 9) filtering of input hyperspectral Raman image, 1D ResUNet output, and target high SNR hyperspectral Raman image with corresponding imaging times shown in white (min:sec) (scale bar = 10 μm). **(d)** Exemplar vertex component analysis (VCA) performed on target high SNR hyperspectral Raman image identifies 5 key components (proteins/lipids [red], nucleic acids [blue], proteins [green], lipids [yellow], and background [black]) which are applied to low SNR input, Savitzky-Golay output, and 1D ResUNet output images via non-negatively constrained least-squares regression demonstrating that low SNR input and Savitzky-Golay output data do not effectively identify different cell components. **(e-f)** Exemplar Raman spectra (white arrows in **(c)**) corresponding to **(e)** a lipid-rich cytoplasmic region and **(f)** the nucleus.

We then examined whether the neural network-based Raman spectral reconstruction would result in significant loss of biochemical information or the introduction of 'hallucinated' spatial or spectral features. To do this, we compared the quality of Raman hyperspectral images for each hyperspectral Raman cell image based on low SNR input data, SG smoothed low SNR data, neural network reconstructed data, and the ground truth high SNR data (Figure 2c). SG filtering produced a spatially blurred image with a loss of fine details, as it is performed independently for each pixel, while the 1D ResUNet produced a much sharper result with improved contrast of biomolecular features that closely aligns with the target high SNR image, suggesting the preservation of biochemical information. Indeed, the 1D ResUNet significantly outperformed SG filtering in terms of two commonly used image quality metrics, peak signal-to-noise ratio (PSNR) and structural similarity index (SSIM). While the 1D ResUNet achieved a mean PSNR of 46.21 [95% CI: 45.76, 46.67] and a mean SSIM of 0.9532 [95% CI: 0.9154, 0.9910] across the 11 hyperspectral Raman cell images, SG filtering resulted in the statistically significantly lower values (assessed via a two-tailed Wilcoxon paired signed rank test) of 38.53 [95% CI: 37.18, 39.88] and 0.8325 [95% CI: 0.7434, 0.9216], respectively.

To assess the preservation of biochemical information, we next applied a vertex component analysis (VCA) to an exemplar target high SNR hyperspectral Raman image (Figure 2d). VCA is a spectral unmixing technique designed for the unsupervised extraction of endmembers from hyperspectral data, enabling the identification of major constituent components in a hyperspectral Raman image (e.g. lipid-rich, nucleic-acid rich, background regions)[38]. The VCA endmembers identified for the target high SNR hyperspectral Raman image (Supplementary Figure 2) were then applied to the input data, the SG output data, and the 1D ResUNet output data via non-negatively constrained least-squares regression. Crucially, this analysis demonstrated that the 1D ResUNet output effectively identified and preserved key molecular species present in the hyperspectral image in line with the target high SNR hyperspectral image. In contrast, the SG output failed to robustly distinguish the different Raman cellular signatures. Exemplar spectra from two different regions (nucleus and cytoplasm) of the hyperspectral images (Figure 2e-f) further demonstrated the superiority of the 1D ResUNet output for the accurate reconstruction of biochemical information contained in the Raman spectra. Importantly, applying the neural network on a per-hyperspectral pixel in this manner effectively enabled Raman spectroscopic imaging up to 10× faster than conventional Raman spectroscopy in this case, while preserving biochemical information.

**Deep Learning Enabled Hyperspectral Image Super-Resolution**

While the denoising results demonstrate a significant improvement in imaging times for conventional Raman spectroscopic imaging, the 1D ResUNet does not consider the high degree of molecular compositional correlation between adjacent pixels. We therefore sought to improve this and take spatial context into consideration by developing a 2D neural network for hyperspectral Raman image super-resolution (HyRISR). To achieve this, we trained HyRISR to take a LR (high SNR) hyperspectral image as input and output a corresponding HR (high SNR) hyperspectral image. HyRISR learns to identify spatial and spectral correlations present in the training set in order to develop an accurate mapping between LR inputs and target HR outputs in the test set. HyRISR follows a similar architecture to the RCAN, with the introduction of spectral downsampling early in the network followed by spectral upsampling at the end of the network (Supplementary Figure 1). This use of spectral downsampling exploits the high spectral resolution (and hence high channel redundancy) of Raman spectra to reduce the computational load of HyRISR without sacrificing super-resolution performance.

We applied HyRISR to a dataset of hyperspectral Raman images of MDA-MB-231 breast cancer cells (n = 169 Raman images) using a data split of 85:10:5 for training, validation, and test sets (see Materials and Methods and Supplementary Table 2 for full implementation details). To increase the effective size of our dataset and improve the robustness of HyRISR, we employed extensive data augmentation including randomly applied image cropping, flipping, rotation, and mixup[39], as well as randomly applied spectral flipping and shifting (Supplementary Figure 3). To generate LR input images, we downsampled corresponding HR images 2×, 3×, or 4× by keeping only every $2^{nd}$, $3^{rd}$, or $4^{th}$ pixel (where each pixel of the hyperspectral image contains a full Raman spectrum) in both $x$ and $y$ to reflect the raster scan nature of Raman imaging. Application of HyRISR to the test set for 2×, 3×, and 4× super-resolution yielded superior performance, statistically significantly exceeding standard nearest neighbour and bicubic upsampling methods in terms of two image quality metrics, PSNR and SSIM, as well as in terms of MSE (Figure 3, Supplementary Figures 4-6). Importantly, these results, particularly the 2× and 3× super-resolution images, demonstrated good fidelity to the corresponding high-resolution target image, accurately reconstructing spectral features to correctly identify cellular components via VCA. This, combined with the reduced MSE values, demonstrates that the neural network can effectively preserve molecular information through accurate spectral reconstruction. Although at 2× HyRISR produced minimal blurring, blur increased significantly at 3× and 4×, a well-known result of training our SISR neural network with an L1 loss function (see Discussion for further details). Despite this blurring, the HyRISR output qualitatively better delineated cell boundaries, correctly identified subcellular features, and introduced fewer artefacts compared to bicubic upsampling.

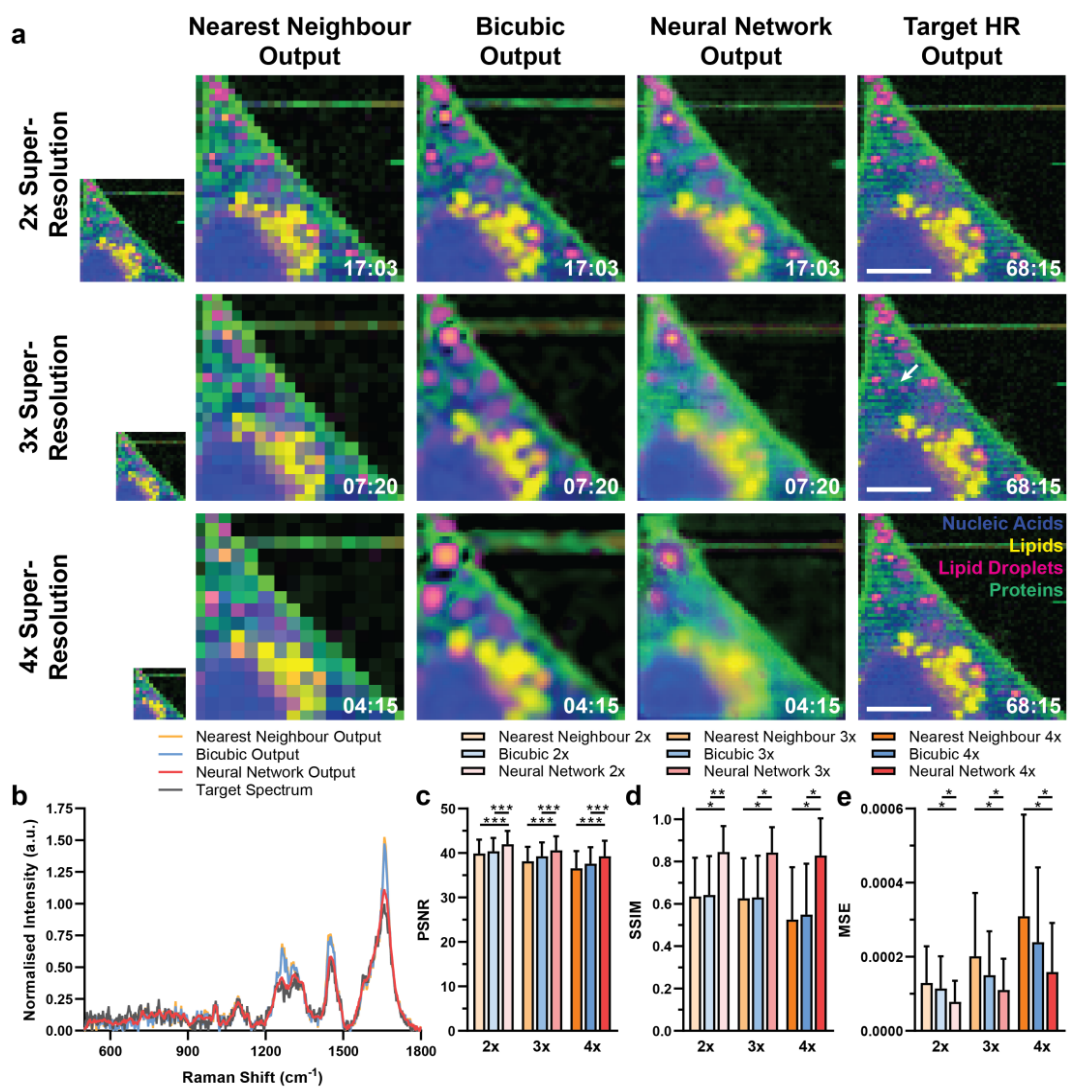

**Figure 3 | Deep learning enabled hyperspectral image super-resolution. (a)** 2×, 3×, and 4× super-resolution of example test set hyperspectral Raman image enables a significant reduction in imaging times (shown in white) while recovering important spatial and spectral information (scale bars = 10 μm). Images shown are the result of a VCA performed on the target HR hyperspectral Raman image, which identified 5 key components (lipid droplets [magenta], nucleic acids [blue], proteins [green], lipids [yellow], and background [black]). VCA components were applied to the nearest neighbour output, bicubic output, and HyRISR output images via non-negatively constrained least-squares regression. **(b)** Exemplar Raman spectrum at white arrow in **(a)** demonstrating that the neural network output (red) is more closely aligned to the target (ground truth) spectrum (dark grey). **(c-d)** Mean test set **(c)** PSNR, **(d)** SSIM, **(e)** MSE values for nearest neighbour upsampling, bicubic upsampling, and HyRISR output for 2×, 3×, and 4× super-resolution (n = 9) (error bars: mean ± STD) (One-way paired analysis of variance (ANOVA) with Geisser-Greenhouse correction and Tukey's multiple comparisons test, * $P < 0.05$, ** $P < 0.01$, *** $P < 0.001$).

Notably, as with the denoising 1D ResUNet presented above, HyRISR enabled a significant reduction in imaging time, down from 68:15 (min:sec) to 17:03 in the case of 2× super-resolution and to 07:20 for 3× super-resolution. Importantly, this was achieved with only a limited loss of high-frequency details, with biochemical spectral information well maintained as evidenced by VCA (Figure 3a-b). In contrast, bicubic upsampling introduced numerous artefacts into the hyperspectral image. Although 4× super-resolution reduces imaging time further to 04:15, it does so with a much greater loss of fine details for both bicubic upsampling and HyRISR. While this might not be suitable for high-resolution cellular imaging, such 4× super-resolution Raman imaging could prove useful for other applications.

**Hybrid Denoising and Super-Resolution Raman Spectroscopy for High-Throughput Molecular Imaging**

While both our 1D ResUNet Raman spectral denoising and HyRISR neural networks enable significant speed-ups in Raman imaging time, single cell Raman imaging using either network remains on the order of minutes. Though this is considerably faster than conventional Raman imaging, it remains too slow for high-throughput applications such as cell imaging or automated spectral histopathology. To further improve the speed of Raman image acquisition, we next combined our two neural networks to perform sequential Raman spectral denoising followed by hyperspectral image super-resolution. Sequential application of the neural networks in this manner enabled the use of all data present in each dataset (as opposed to the small subset of data present in both the denoising and HyRISR datasets). Here, using the Raman spectra from a single cell in the test set for both datasets, we achieved effective speed-ups of 40× (2× super-resolution), 90× (3× super-resolution), and 160× (4× super-resolution) while accurately reconstructing a high SNR, HR hyperspectral Raman image from a low SNR, LR input hyperspectral Raman image (Figure 4, Supplementary Figure 7).

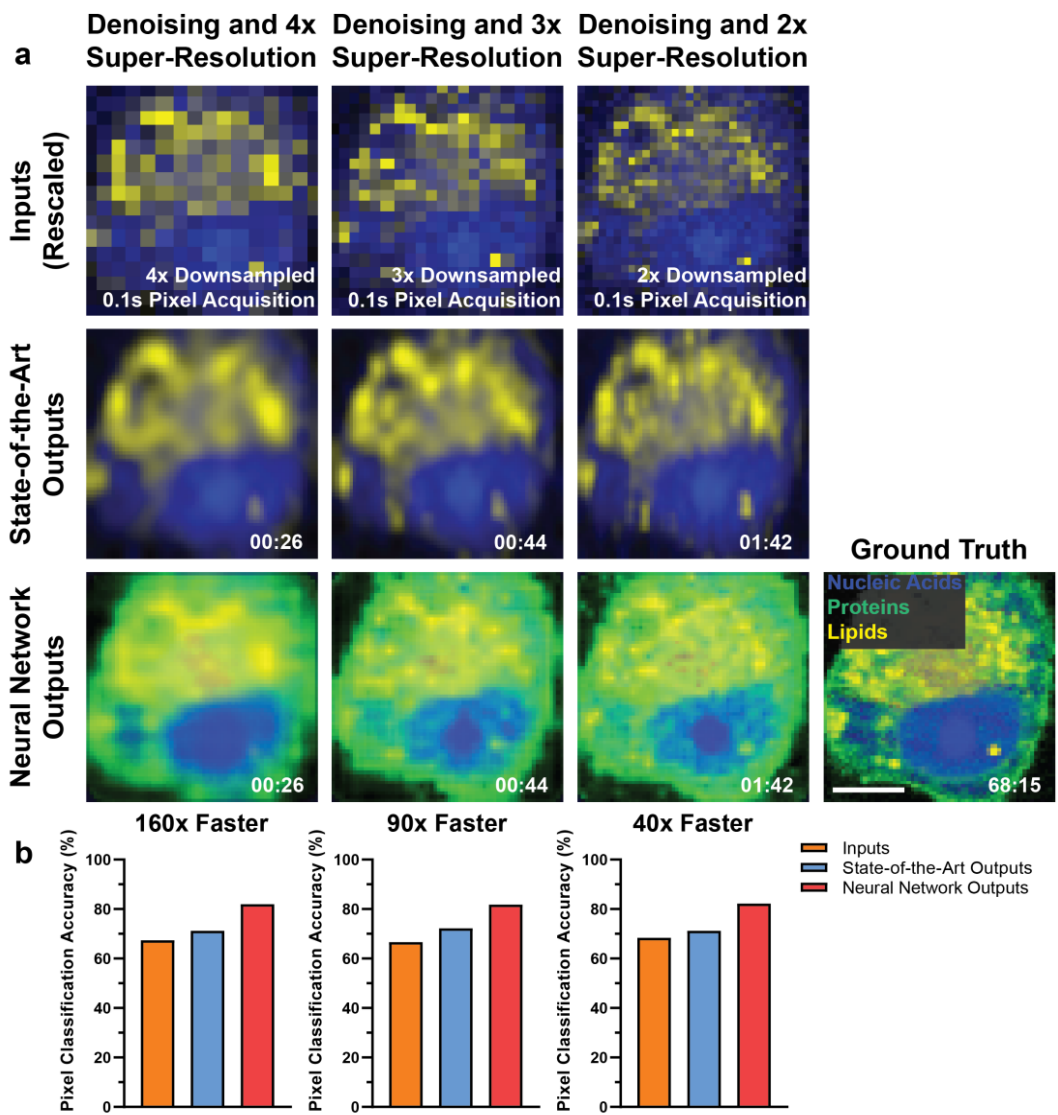

**Figure 4 | Combined Raman spectral denoising and hyperspectral image super-resolution enables extreme speed-ups in Raman imaging time. (a)** Sequential application of Raman spectral denoising followed by hyperspectral image super-resolution enables extreme speed-ups in imaging time (shown in white) from 68:15 (min:sec) to 01:42 for 2× super-resolution, 00:44 for 3× super-resolution, and 00:26 for 4× super-resolution while largely preserving molecular information (scale bars = 10 μm). Images shown are the result of a VCA performed on the target HR, high SNR hyperspectral Raman image, which identified 4 key components (nucleic acids [blue], proteins [green], lipids [yellow], and background [black]). VCA components were applied to input, state-of-the-art output (SG filtering and bicubic upsampling), and neural network output images via non-negatively constrained least-squares regression. **(b)** Pixel classification accuracy for input, state-of-the-art output (SG filtering and bicubic upsampling), and neural network output images as compared to VCA pixel classification of target HR, high SNR hyperspectral Raman image.

We again used VCA to identify key Raman spectral components in our ground truth image and employed non-negatively constrained least-squares regression to apply the identified VCA endmembers to the input images (nearest neighbour rescaled), state-of-the-art output images (SG filtering and bicubic upsampling), and neural network output images. Our neural networks outperformed the combination of SG filtering and bicubic upsampling, accurately reconstructing both spatial and spectral information and maintaining accurate VCA endmember identification (Figure 4a). In each case, this resulted in an improved pixel classification accuracy (as compared to pixel classification for the ground truth hyperspectral Raman image, determined as the VCA endmember with the maximum intensity value for each pixel as per non-negatively constrained least-squares regression) relative to the inputs and state-of-the-art outputs (Figure 4b). Crucially, accurate spectral and spatial reconstruction was maintained even for 40× and 90× Raman imaging time speed-ups, enabling HR hyperspectral Raman cell imaging in under 2 minutes or under 1 minute, respectively. While imaging time can be further reduced by employing 4× super-resolution for a 160× Raman imaging time speed-up, reconstructed image quality continues to degrade and may produce undesirable artefacts at higher super-resolution scales.

**Generalised Hyperspectral Imaging from Cells to Tissues Using Transfer Learning**

Finally, to highlight the generalisability and wide applicability of our DeepeR framework, we used transfer learning to apply HyRISR to a small dataset of hyperspectral Raman images of unrelated origin. We thus extended DeepeR to the field of regenerative medicine for super-resolution hyperspectral Raman imaging of *in vitro*-formed cartilage constructs. Both the spatial and spectral information contained in the hyperspectral Raman images of tissue-engineered cartilage samples differ significantly from those of MDA-MB-231 breast cancer cells used to train HyRISR, representing an effective test of the transferability of DeepeR. While we have already demonstrated the potential of DeepeR for hyperspectral Raman cellular image super-resolution, deep learning is a data-heavy approach that requires large, labelled datasets in order to be effective. For applications where such a large dataset does not exist, data acquisition for deep learning can be prohibitively time-consuming and expensive. Here, we aimed to demonstrate that transfer learning, the application of an existing neural network model trained on a large dataset to a second, smaller dataset, can achieve high quality results (Figure 1 iv).

To do this, we used a small training dataset consisting of 16 patches (64×64 pixels each) from large HR Raman hyperspectral images of tissue-engineered cartilage, with a separate test set of 12 patches (64×64 pixels each) from a single tissue-engineered cartilage sample. Here, transfer learning was performed by further training all neural network weights from a pre-trained HyRISR model for 200 epochs on the small tissue-engineered cartilage training dataset. We then compared the super-resolution performance of this fine-tuned model, against both nearest neighbour and bicubic upsampling as well as against HyRISR trained from scratch on the small training dataset of hyperspectral Raman tissue-engineered cartilage image patches alone (Figure 5, Supplementary Figure 8). As expected, transfer learning of HyRISR achieved superior results to nearest neighbour upsampling, bicubic upsampling, and HyRISR trained from scratch on the tissue-engineered cartilage dataset alone in terms of PSNR and SSIM (Figure 5c-d). As with super-resolution of hyperspectral Raman images of MDA-MB-231 breast cancer cells, the fine-tuned neural network here produced a highly accurate reconstruction with few introduced artefacts and a degree of over-smoothing. Meanwhile, bicubic upsampling resulted in an image that appears grossly similar to the target ground truth image but suffered from the introduction of numerous artefacts, resulting in both spatial and spectral distortion (Figure 5b-d).

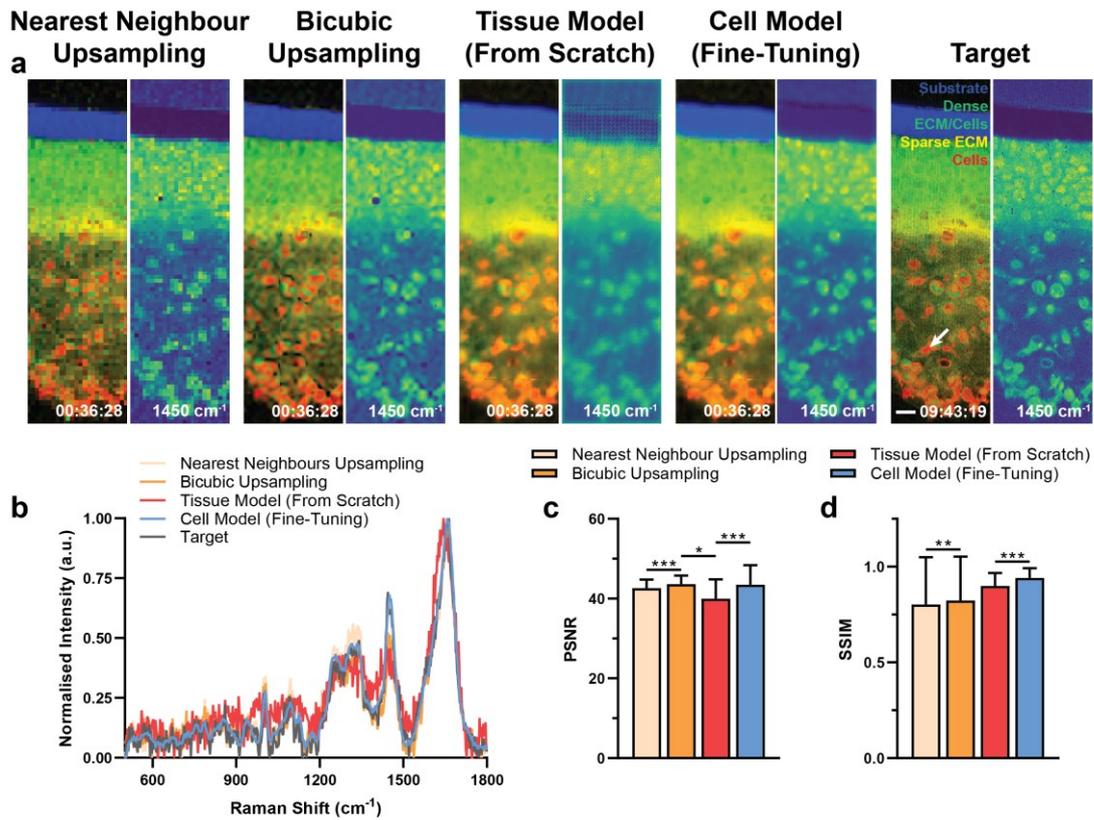

**Figure 5 | Transfer learning enables effective super-resolution for a small dataset of tissue-engineered cartilage hyperspectral Raman images. (a)** Transfer learning of our HISR neural network, trained only on MDA-MB-231 breast cancer cell images, enabled effective cross-domain 4× super-resolution of hyperspectral Raman images despite having only a very small dataset of tissue-engineered cartilage for training. For each condition, images shown on the left are the result of a VCA performed on the target HR, high SNR hyperspectral Raman image, which identified 5 key components (substrate [blue], dense ECM/cells [green], sparse ECM [yellow], cells [red], and background [black]). VCA components were applied to nearest neighbour upsampling, bicubic upsampling, tissue model (from scratch), and cell model (fine-tuning) images via non-negatively constrained least-squares regression. Images shown on the right for each condition are 1450 cm$^{-1}$ peak intensity heatmaps. All images formed as composition of overlapping 64x64 pixel image patches (scale bars = 10 μm). **(b)** Exemplar Raman spectrum (white arrow in **(a)**) demonstrating that transfer learning achieves high accuracy reconstruction of the target spectra for each pixel. **(c-d)** Mean test set **(c)** PSNR and **(d)** SSIM values for nearest neighbour upsampling, bicubic upsampling, and neural network outputs for 4x super-resolution, calculated on a per-image patch basis (n = 12 patches) (Error bars: mean ± STD) (One-way paired analysis of variance (ANOVA) with Geisser-Greenhouse correction and Tukey's multiple comparisons test, * $P < 0.05$, ** $P < 0.01$, *** $P < 0.001$).

## Discussion

DeepeR is a comprehensive deep learning framework that offers a completely new approach to high-throughput Raman spectroscopic imaging. We first demonstrated that deep learning could be used to reconstruct high SNR Raman spectra from low SNR inputs, effectively preserving crucial biochemical information whilst removing unwanted noisy artefacts (Figure 2). Such a spectral-based approach could be applied to either spectroscopic imaging (as shown here) or point spectroscopy, enabling a significant reduction in spectral acquisition times. Our neural network approach vastly outperformed the current state-of-the-art in Raman filtering, more accurately preserving biochemical information while enabling a reduction in hyperspectral Raman imaging time of up to 10×.

We next developed a neural network for super-resolution of hyperspectral Raman images, designed to reconstruct both spatial and spectral information and deliver a substantial reduction in imaging time. Our neural network outperformed nearest neighbour and bicubic upsampling methods in terms of PSNR, SSIM, and MSE, and achieved speed-ups of 4×, 9×, or 16× depending on the super-resolution scale (Figure 3).

We then sought to achieve extreme speed-ups in hyperspectral Raman imaging times through the combination of our denoising and super-resolution neural networks for high-throughput Raman imaging (Figure 4). In doing so, we were able to produce HR, high SNR hyperspectral Raman images from LR, low SNR inputs and achieve speed-ups of up to 160× while maintaining high spectral and spatial fidelity to the target ground truth HR, high SNR hyperspectral Raman image as evidenced by regression of VCA endmembers. This approach enabled a reduction in Raman cell imaging time from over 68 minutes to less than 1 minute.

Finally, we demonstrated the generalisability of DeepeR for hyperspectral Raman imaging, showing how transfer learning could be used to extend pre-trained models to an unrelated biomedical application (Figure 5). DeepeR can be applied online or offline to existing Raman spectroscopic systems without requiring any hardware modifications or imposing system limitations using transfer learning. Offline application to existing hyperspectral Raman datasets could be used to develop custom models for specific applications, either from scratch or by transfer learning from our pre-trained networks. Online application, with inference occurring in a matter of seconds for a GPU-equipped scientific computer, will deliver high-throughput imaging capabilities, transforming the potential range of applications for hyperspectral Raman imaging.

DeepeR will thus help to drive forward high-throughput hyperspectral Raman imaging, representing a major departure from existing chemometric and other multivariate statistical techniques. DeepeR achieves this by effectively learning the complex spectral and spatial correlations that exist in our large dataset of hyperspectral Raman images and exploiting these to achieve state-of-the-art results in Raman spectral denoising and hyperspectral Raman image super-resolution. Importantly, beyond denoising and super-resolution applications, our results demonstrate the potential of deep learning to improve hyperspectral Raman analysis with implications for clinical diagnostics, chemometric analysis, and sample identification among others.

Despite its significant advances, DeepeR does face a number of limitations that must be considered before application to additional hyperspectral Raman datasets. Most notably, our framework is unlikely to accurately reconstruct very fine (e.g. single pixel) details and so may not be suitable for HR imaging of small, complex specimens. This is analogous to the difficulties SISR neural networks face in reconstructing intricate features present in RGB images[40]. In contrast, the sparsity of such fine details in larger specimens such as tissues makes them well suited to our deep learning framework. Secondly, while transfer learning using our pre-trained models will enable a much wider range of applications, hyperspectral Raman images with substantially different spatial and spectral features will still require a sufficiently large dataset for effective performance. Future work will seek to expand the depth and breadth of our hyperspectral Raman dataset, encompassing spectra from a variety of instruments and samples. This will increase both the performance and generalisability of DeepeR. Lastly, before widespread application to Raman spectroscopic imaging is possible, large-scale prospective validation will need to be performed, specific to each application, to ensure diagnostic or scientific decisions match those made for corresponding HR, high SNR hyperspectral Raman data.

There remains scope for improvements in the performance of our deep learning framework, most notably through the collection of larger training datasets from different biomedical applications and the development of more advanced neural network architectures. Collection of a large dataset of paired low SNR, LR and high SNR, HR hyperspectral Raman images would enable the training of a joint denoising and super-resolution neural network, which we anticipate would produce improved performance in line with existing studies on multi-task neural networks[41]. Similarly, performance could be further improved by implementing a generative adversarial network (GAN) architecture[42]. GANs have demonstrated an array of impressive results for super-resolution of RGB images and medical images such as endomicroscopy images, outperforming networks trained with an L1 loss function in terms of human visual perception, and would likely achieve similar results for hyperspectral images[24,43,44]. However, GAN architectures pose particularly significant demands on computational resources in the context of hyperspectral image super-resolution.

Though here we demonstrated application of DeepeR to Raman spectroscopy, equivalent neural network architectures could be generalised to alternative techniques such as FT-IR imaging[45], hyperspectral imaging[46], or mass spectrometry imaging techniques[47]. DeepeR also paves the way for future pragmatic developments such as robust autofluorescence removal, segmentation or even automated label-free staining across multiple spectral imaging techniques.

In conclusion, DeepeR represents a comprehensive deep learning framework for high-throughput hyperspectral Raman imaging. This has the potential to transform the application of Raman spectroscopic imaging in the biomedical sciences, enabling a host of higher-throughput applications not previously possible. Crucially, the information and data we provide open source to the community, including our complete dataset, pre-trained models, and Python code will enable rapid expansion and integration of our framework into existing Raman spectroscopy systems, driving forward high-throughput Raman imaging.

## Materials and Methods
### Cell Culture
MDA-MB-231 breast cancer cells were originally obtained from the ATCC (Manassas, VA, USA) and authenticated via STR profiling. Cells were maintained in Dulbecco's modified Eagle's medium (DMEM), supplemented with 1x non-essential amino acids, 25 mM Hepes, 1x penicillin/streptomycin, and 10% (v/v) foetal bovine serum (FBS), all obtained through Gibco (Thermo Fisher Scientific, Inc., Waltham, MA, USA). Cells were grown on magnesium fluoride ($MgF_2$) windows (Global Optics Ltd, Bournemouth, UK). 20,000 cells per $cm^2$ were seeded in serum-supplemented DMEM and allowed to adhere overnight. Prior to imaging, cells were washed with DPBS and fixed with 4% (v/v) paraformaldehyde (PFA, Sigma-Aldrich, St Louis, MO, USA) in DPBS for 30 minutes at room temperature. PFA solution was aspirated and samples washed three times with DPBS.

### Tissue Culture and Sample Preparation
Articular cartilage was excised aseptically from the full thickness of metacarpal−phalangeal joints of mature cows (n = 3) aged 24 to 36 months within 48 h of death. Chondrocytes were isolated from the tissue by sequential enzymatic digestion at 37 °C (0.2% (w/v) Pronase (Roche Applied Science) in DMEM with 4.5 g/L glucose (Invitrogen); for 1 h followed by 0.04% (w/v) collagenase type I (Sigma-Aldrich) in DMEM overnight. Hydrophilic polytetrafluoroethylene (PTFE) membranes (Millipore) encased in a cell chamber were incubated with 100 μL of 0.5 mg/mL collagen type II (Sigma-Aldrich) in 0.1 N acetic acid overnight to allow full evaporation of the solution and then washed 3 times in PBS. Cell chambers were placed in 24 well plates, and 750 μL of DMEM supplemented with 5% (v/v) FBS (HyClone) was added to each well outside the cell chamber. Isolated chondrocytes were seeded on top of PTFE membrane inserts (1 × $10^6$ cells in 750 μL per membrane; 12 mm diameter) in either 2 mL of DMEM with 1 g/L glucose, 11 mL of DMEM with 1 g/L glucose, or 11 mL of DMEM with 4.5 g/L glucose; supplemented with 5% (v/v) FBS and incubated at 37 °C and 5% $CO_2$. On day 2, the medium was supplemented with ascorbic acid (50 μg/mL; Sigma-Aldrich) and the FBS supplementation was increased to 10% (v/v). Culture medium was changed every 2−3 days and cultures harvested at 14, 28, or 42 days. Tissue-engineered cartilage constructs were washed 3 times for 15 min in PBS, and fixed in 4% (v/v) paraformaldehyde for 30 min.

Tissue-engineered cartilage constructs were cemented onto a polystyrene surface using a small quantity of cyanoacrylate. These were then immersed in a drop of water and frozen on a cryosectioning block. Using a cryostat (Bright Instruments Ltd.), the tissues were sectioned flat at a 45° angle relative to the articular surface. Samples were stored at 4 °C in PBS until Raman spectroscopic imaging was performed. Tissue cross-section samples imaged by Raman spectroscopy were prepared to a thickness of greater than 1 mm.

### Raman Spectroscopic Imaging
Raman imaging was performed using a confocal Raman microscope (alpha300R+, WITec, GmbH, Germany). A 532 nm laser light source at 35 mW power output was applied through a 63x/1.0 NA water-immersion microscope objective lens (W Plan-Apochromat, Zeiss, Germany). Inelastically-scattered light was collected through the objective and directed via a 100 μm diameter silica fibre, acting as a confocal pinhole, to a high-throughput imaging spectrograph (UHTS 300, WITec, GmbH, Germany) with a 600 groove/mm grating and equipped with a thermo-electrically cooled (-60 °C) back-illuminated charge-coupled device (CCD) camera. Hyperspectral Raman images were acquired with a 0.5 μm spatial resolution (1 μm spatial resolution for tissue-engineered cartilage) and a 1 second integration time, where for each pixel a Raman spectrum was acquired in the range from 0 to 3700 $cm^{-1}$ with a spectral resolution of 11 $cm^{-1}$.

### Hyperspectral Raman Image Processing
Preliminary hyperspectral Raman image processing was kept constant for all datasets and performed using Witec ProjectFOUR software. Spectra were cropped to the fingerprint spectral region (600 – 1800 $cm^{-1}$) and background autofluorescence subtraction performed using ProjectFOUR's 'shape' background filter with α = 500. Hyperspectral images were then exported to MATLAB.

### Neural Network Architecture and Implementation: Raman Spectral Denoising
Denoising of Raman spectra was achieved via a 1D ResUNet architecture (Figure 1). The network was trained for 500 epochs using the Adam optimizer[48], with an L1-norm loss function and a one-cycle learning rate scheduler. 11 independent models were trained, with all of the Raman spectra from a single hyperspectral Raman cell image used as the test set in each case. In each case, the training and validation sets were formed from all of the Raman spectra from the remaining 10 hyperspectral Raman cell images. Evaluation on the validation set was used to prevent overfitting, while all results presented are the mean across the 11 test set folds. Full training details are provided in Supplementary Table 1. See Supporting Information for complete Python training scripts.

**Neural Network Architecture and Implementation: Hyperspectral Super-Resolution**
Hyperspectral Raman image super-resolution was performed using a hyperspectral residual channel attention network (Supplementary Figure 1). The network was trained for 600 epochs using the Adam optimizer[48], with an L1-norm loss function and a constant learning rate. Evaluation on the validation set was used to prevent overfitting, while all results presented are for the test set. Full training details are provided in Supplementary Table 2. See Supporting Information for complete Python training scripts. Network performance was assessed using two common image quality metrics, peak signal-to-noise ratio (PSNR) and structural similarity index (SSIM).

The PSNR of an image, $y$, relative to an uncorrupted ground truth image, $x$, is defined as:

$$PSNR(x, y) = 10 \cdot \log_{10}\left(\frac{x_{max}^2}{MSE}\right)$$

Where $x_{max}$ is the maximum possible pixel value (across all channels) of the ground truth image, $x$, and the mean squared error (MSE), calculated across all channels, is defined as:

$$MSE = \frac{1}{m \cdot n \cdot p} \sum_{i=0}^{m-1} \sum_{j=0}^{n-1} \sum_{k=0}^{p-1} [x(i,j,k) - y(i,j,k)]^2$$

The SSIM between two images, $x$ and $y$, is a defined as

$$SSIM(x, y) = \frac{(2\mu_x \mu_y + c_1)(2\sigma_{xy} + c_2)}{(\mu_x^2 + \mu_y^2 + c_1)(\sigma_x^2 + \sigma_y^2 + c_2)}$$

Where $\mu_x$ and $\mu_y$ are the means of $x$ and $y$, $\sigma_x^2$ and $\sigma_y^2$ are the variances of $x$ and $y$, and $c_1$ and $c_2$ are two numerical constants to stabilize the division. For hyperspectral Raman images, the SSIM is calculated independently for each channel and then averaged.

**Data Augmentation**
Data augmentation, essential for increasing the effective dataset size, was performed using a custom PyTorch DataGenerator. Data augmentation included image subsampling, flipping, rotation, and mixup[39], as well as spectral shifting and flipping (Supplementary Figure 3).

**Implementation**
Complete implementation details including dataset size, training/validation/testing splits, hyperparameter selection, and training times are shown in Supplementary Tables 1 and 2.

Network training and optimisation was performed using the Joint Academic Data science Endeavour (JADE) high performance computing (HPC) facility on an NVIDIA DGX-1 deep learning system with 8 NVIDIA V100 GPUs.

Final network implementation and training was performed in Python 3.7.3 using PyTorch 1.4.0 on a desktop computer with a Core i7-8700 CPU at 3.2 GHz (Intel), 32 GB of RAM, and a Titan V GPU (NVIDIA), running Windows 10 (Microsoft).

**Supporting Information**
Datasets will be made available upon publication. All code and pretrained models are available online at https://github.com/conor-horgan/.

**Acknowledgements**
This work has received funding from the European Research Council (ERC) under the European Union's Horizon 2020 research and innovation programme (grant agreement No. 802778). This grant has allocated funding for open access publication. This work is supported by the Wellcome/EPSRC Centre for Medical Engineering [WT 203148/Z/16/Z]. CCH acknowledges funding from the NanoMed Marie Skłodowska-Curie ITN from the H2020 programme under grant number 676137. TV is supported by a Medtronic/Royal Academy of Engineering Research Chair [RCSRF1819/7/34]. AN and MMS acknowledge support from the GlaxoSmithKline Engineered Medicines Laboratory. M.M.S. acknowledges a Wellcome Trust Senior Investigator Award (098411/Z/12/Z). We acknowledge use of the JADE HPC facility, which has received funding through the Engineering and Physical Sciences Research Council (EPSRC).

**Conflict of Interest**
TV is founding director and shareholder of Hypervision Surgical Ltd and holds shares from Mauna Kea Technologies. The authors declare no other competing financial interest.

# Supplementary Information

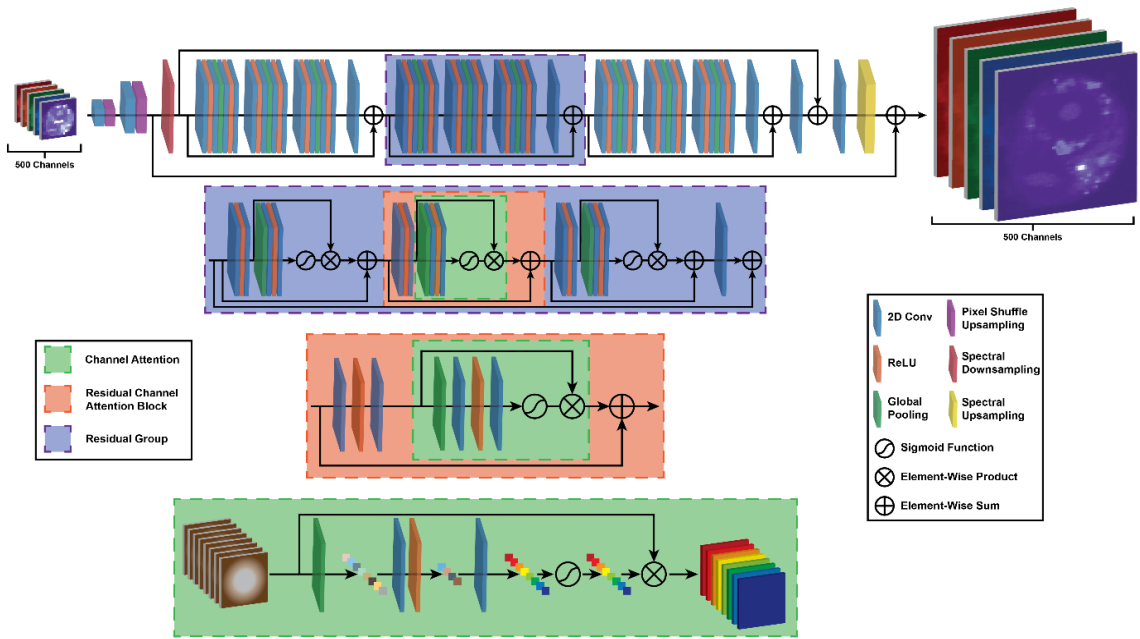

**Supplementary Figure 1 | Hyperspectral Residual Channel Attention Network.** Hyperspectral residual channel attention network architecture employed for super-resolution of hyperspectral Raman images. The network consists of multiple repeated sub-units including the residual groups (purple), residual channel attention blocks (orange), and the channel attention blocks (green). In the final network architecture, *n*(residual groups) = 18 and *n*(residual channel attention blocks) = 16. Spectral downsampling is used at the beginning of the network, exploiting the redundancy of information contained in the high spectral resolution Raman spectra, to reduce computational load, with spectral upsampling placed at the end of the network to return a hyperspectral image with the sample spectral resolution and length as the inputs.

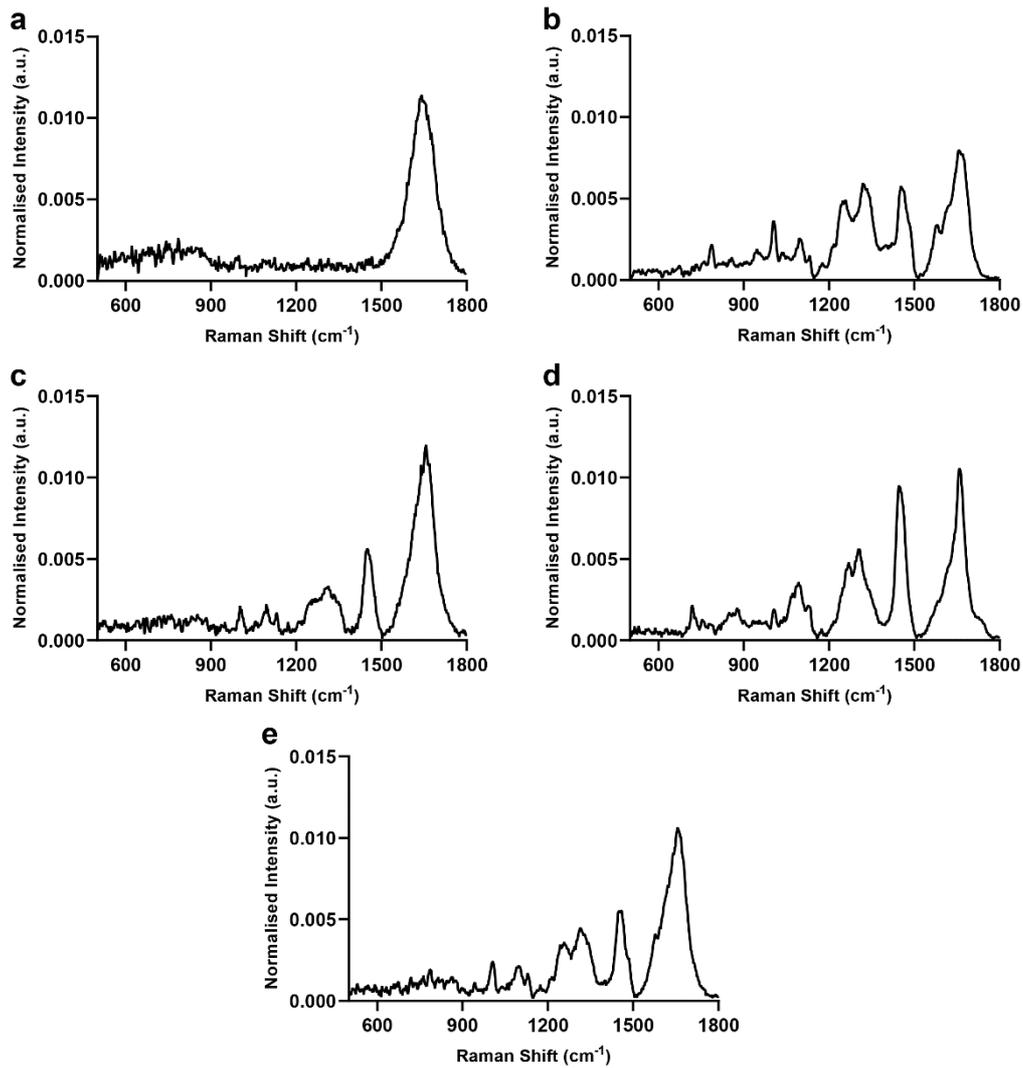

**Supplementary Figure 2 | VCA Endmembers for HR, High SNR Hyperspectral Raman Cell Image for Raman Spectral Denoising. (a-e)** VCA endmembers for high SNR hyperspectral Raman cell image (Figure 2d) corresponding to **(a)** background (black), **(b)** nucleic acids (blue), **(c)** proteins/lipids (red), **(d)** lipids (yellow), and **(e)** proteins (green).

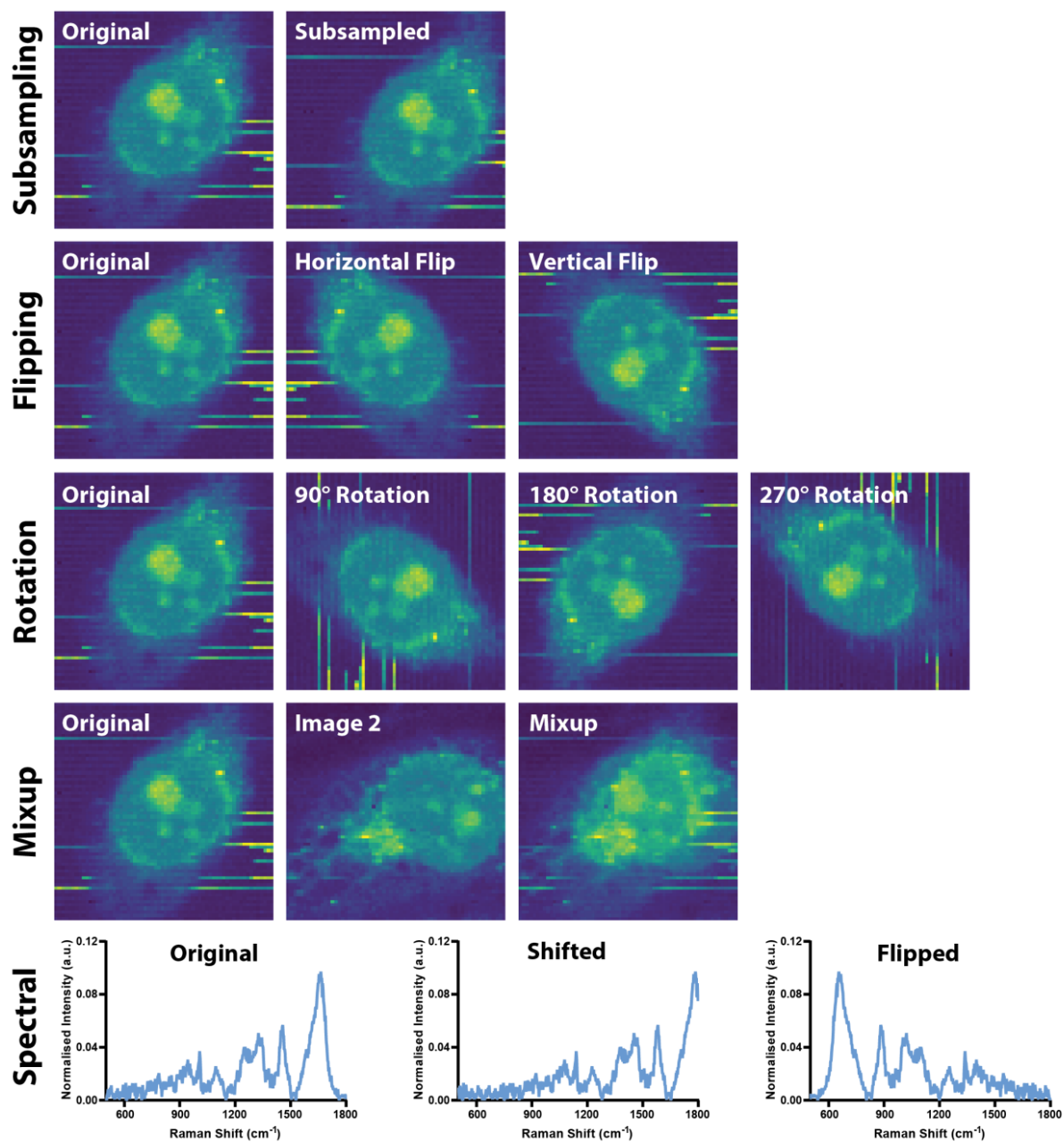

**Supplementary Figure 3 | Data Augmentation for Training of Residual Channel Attention Network for Hyperspectral Image Super-Resolution.** Data augmentations randomly applied to training set of hyperspectral Raman image dataset to increase effective dataset size for effective neural network training. Image data augmentations include subsampling (selecting different sub-regions of an image), horizontal and vertical flipping, rotations, and mixup (the addition of two images following a beta distribution). Additional spectral augmentations include spectral shifting and flipping.

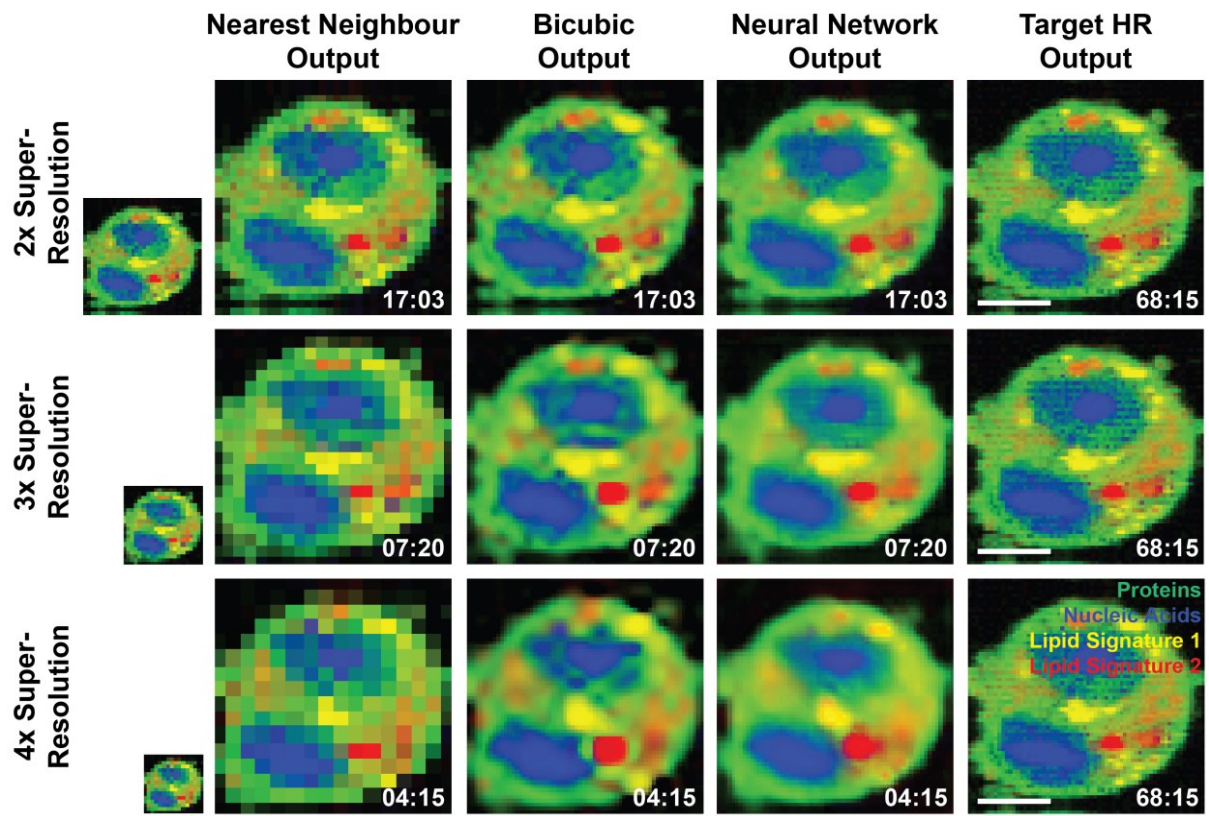

**Supplementary Figure 4 | Deep learning enabled hyperspectral image super-resolution.** 2×, 3×, and 4× super-resolution of example test set hyperspectral Raman image enables a significant reduction in imaging times (shown in white) while recovering important spatial and spectral information (scale bars = 10 μm). Images shown are the result of a VCA performed on the target HR hyperspectral Raman image, which identified 5 key components (proteins [green], nucleic acids [blue], lipid signature 1 [yellow], lipid signature 2 [red], background [black]). VCA components were applied to the nearest neighbour output, bicubic output, and neural network output images via non-negatively constrained least-squares regression.

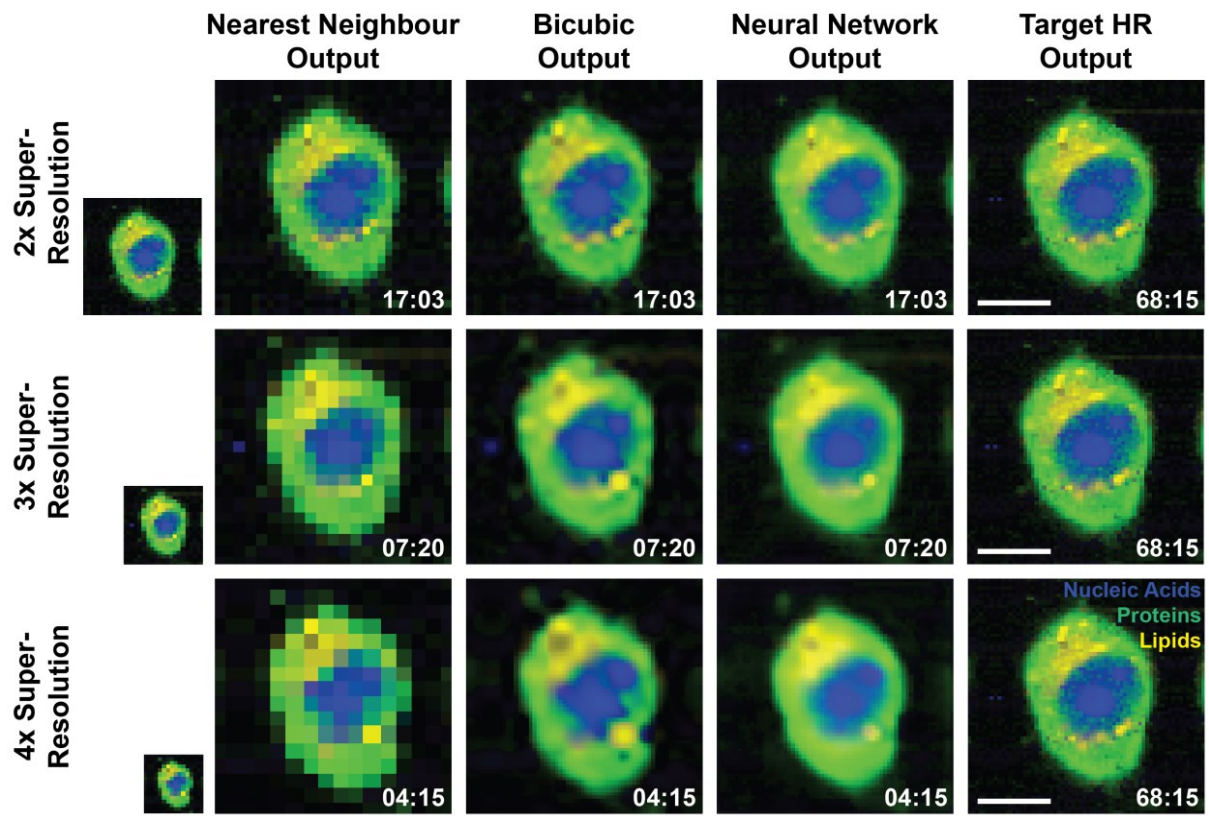

**Supplementary Figure 5 | Deep learning enabled hyperspectral image super-resolution.** 2×, 3×, and 4× super-resolution of example test set hyperspectral Raman image enables a significant reduction in imaging times (shown in white) while recovering important spatial and spectral information (scale bars = 10 μm). Images shown are the result of a VCA performed on the target HR hyperspectral Raman image, which identified 4 key components (nucleic acids [blue], proteins [green], lipids [yellow], background [black]). VCA components were applied to the nearest neighbour output, bicubic output, and neural network output images via non-negatively constrained least-squares regression.

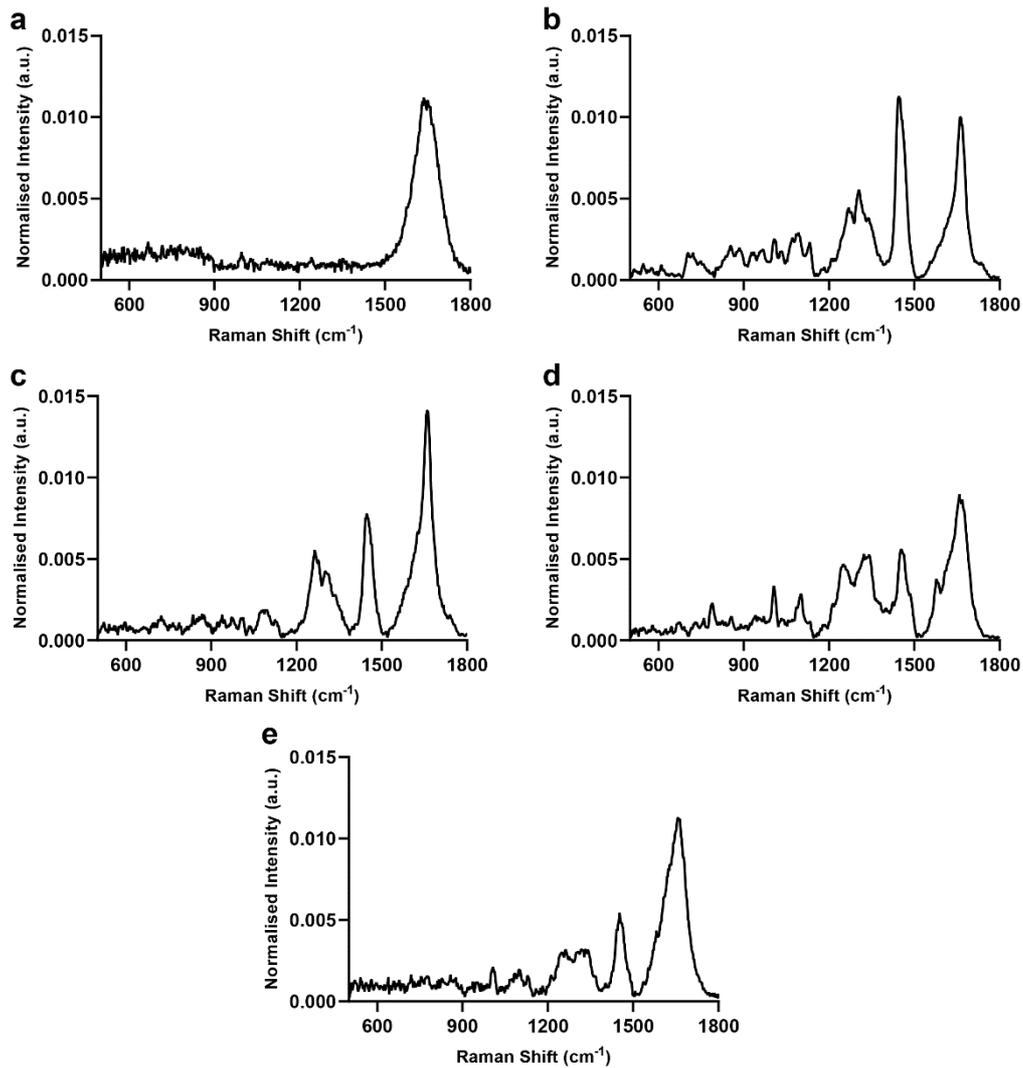

**Supplementary Figure 6 | VCA Endmembers for HR, High SNR Hyperspectral Raman Cell Image for Hyperspectral Image Super-Resolution. (a-e)** VCA endmembers for high SNR hyperspectral Raman cell image (Figure 3) corresponding to **(a)** background (black), **(b)** lipids (yellow), **(c)** lipid droplets (magenta), **(d)** nucleic acids (blue), and **(e)** proteins (green).

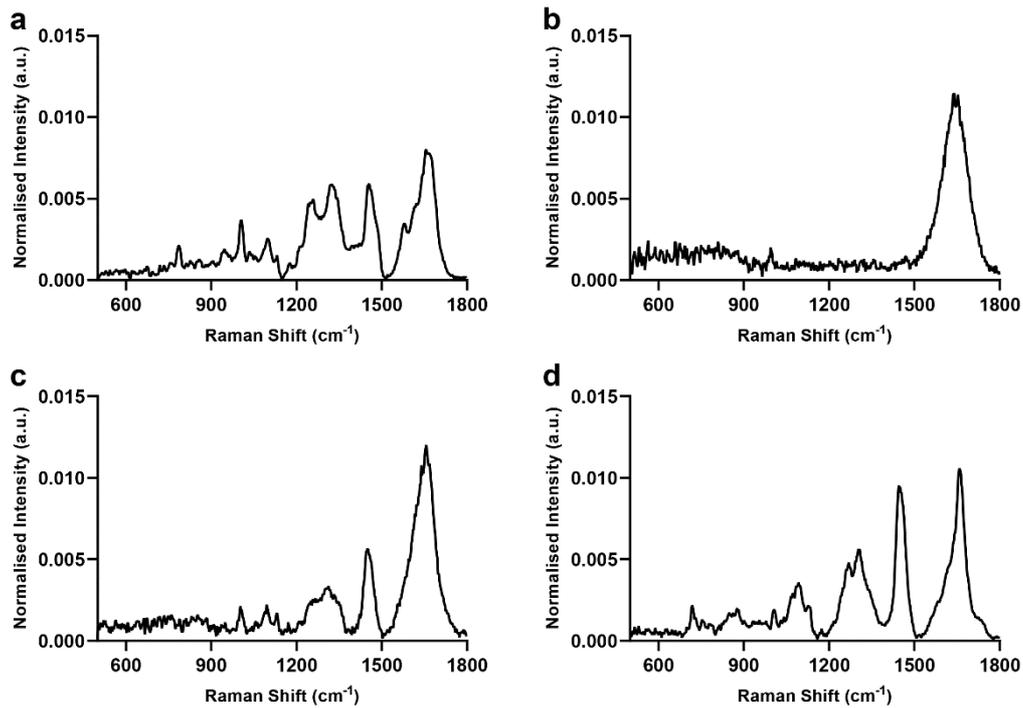

**Supplementary Figure 7 | VCA Endmembers for HR, High SNR Hyperspectral Raman Cell Image for Combined Raman Spectral Denoising and Hyperspectral Image Super-Resolution. (a-d)** VCA endmembers for high SNR hyperspectral Raman cell image (Figure 4) corresponding to **(a)** nucleic acids (blue), **(b)** background (black), **(c)** proteins (green), and **(d)** lipids (yellow).

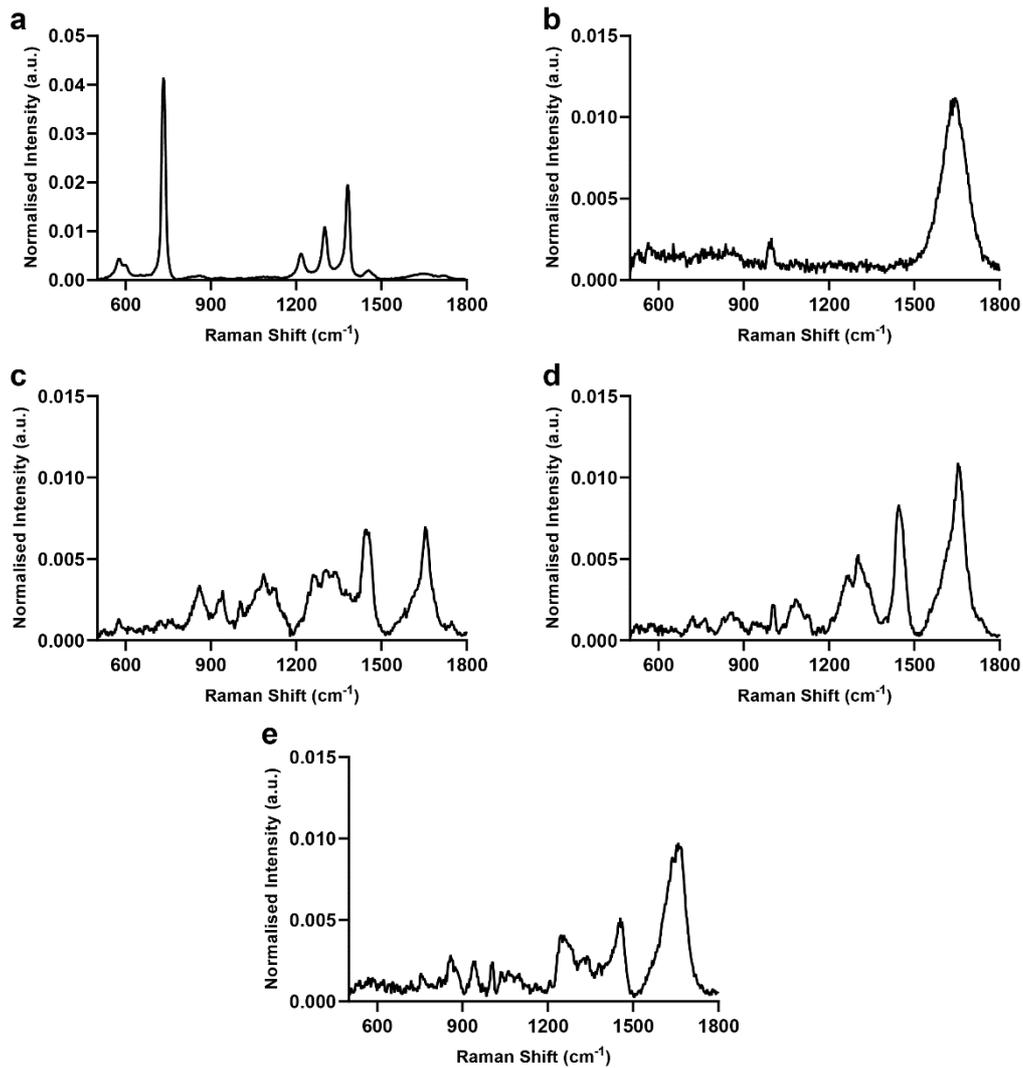

**Supplementary Figure 8 | VCA Endmembers for HR, High SNR Hyperspectral Raman Image of Tissue-Engineered Cartilage for Transfer Learning. (a-e)** VCA endmembers for high SNR hyperspectral Raman cell image (Figure 5) corresponding to **(a)** substrate (blue), **(b)** background (black), **(c)** dense extracellular matrix region (green), **(d)** cells (red), and **(e)** sparse extracellular matrix region (yellow).

**Supplementary Table 1 | Hyperparameter Details for Training of 1D Residual UNet for Raman Spectral Denoising.**

| Hyperparameter | Details |
| --- | --- |
| Dataset size | 172,312 Raman spectra from 11 hyperspectral Raman images |
| Training/validation/testing split | 11-fold leave one image out cross-validation Training/validation: Raman spectra from 10 images (90:10 split) Testing: Raman spectra from 1 remaining image |
| Optimizer | Adam |
| Maximum learning rate | $5\times10^{-4}$ |
| Scheduler | One Cycle LR |
| Epochs | 500 |
| Batch size | 256 |
| Batch norm | Yes |
| Dropout | None |

**Supplementary Table 2 | Hyperparameter Details for Training of Residual Channel Attention Network for Hyperspectral Image Super-Resolution.**

| Hyperparameter | Details |
| --- | --- |
| Dataset size | 169 hyperspectral Raman images |
| Training/validation/testing split | 85:10:5 |
| No. Residual groups | 18 |
| No. Residual blocks | 16 |
| Optimizer | Adam |
| Maximum learning rate | $1\times10^{-5}$ |
| Scheduler | Constant LR |
| Epochs | 600 |
| Batch size | 2 |
| Batch norm | No |
| Dropout | None |